\documentstyle[12pt]{article}

\catcode`\@=11
\long\def\@makefntext#1{
\protect\noindent \hbox to 3.2pt {\hskip-.9pt
$^{{\ninerm\@thefnmark}}$\hfil}#1\hfill}                

 \def\@makefnmark{\hbox to 0pt{$^{\@thefnmark}$\hss}}  
 
\def\ps@myheadings{\let\@mkboth\@gobbletwo
\def\@oddhead{\hbox{}
\rightmark\hfil\ninerm\thepage}
\def\@oddfoot{}\def\@evenhead{\ninerm\thepage\hfil
\leftmark\hbox{}}\def\@evenfoot{}
\def\sectionmark##1{}\def\subsectionmark##1{}}
 
\newcounter{sectionc}\newcounter{subsectionc}\newcounter{subsubsectionc}
\renewcommand{\section}[1] {\vspace{0.6cm}\addtocounter{sectionc}{1}
\setcounter{subsectionc}{0}\setcounter{subsubsectionc}{0}\noindent
	{\bf\thesectionc. #1}\par\vspace{0.4cm}}
\renewcommand{\subsection}[1] {\vspace{0.6cm}\addtocounter{subsectionc}{1}
	\setcounter{subsubsectionc}{0}\noindent
	{\it\thesectionc.\thesubsectionc. #1}\par\vspace{0.4cm}}
\renewcommand{\subsubsection}[1] {\vspace{0.6cm}\addtocounter{subsubsectionc}{1}
	\noindent {\rm\thesectionc.\thesubsectionc.\thesubsubsectionc.
	#1}\par\vspace{0.4cm}}

\newcounter{appendixc}
\newcounter{subappendixc}[appendixc]
\newcounter{subsubappendixc}[subappendixc]

\renewcommand{\appendix}[1] {\vspace{0.6cm}
	\refstepcounter{appendixc}
	\setcounter{figure}{0}
	\setcounter{table}{0}
	\setcounter{equation}{0}
	\renewcommand{\thefigure}{\Alph{appendixc}.\arabic{figure}}
	\renewcommand{\thetable}{\Alph{appendixc}.\arabic{table}}
	\renewcommand{\theappendixc}{\Alph{appendixc}}
	\renewcommand{\theequation}{\Alph{appendixc}.\arabic{equation}}
	\noindent{\bf Appendix \theappendixc #1}\par\vspace{0.4cm}}

\renewenvironment{thebibliography}[1]
	{\begin{list}{\arabic{enumi}.}
	{\usecounter{enumi}\setlength{\parsep}{0pt}
\setlength{\leftmargin 1.25cm}{\rightmargin 0pt}
	 \setlength{\itemsep}{0pt} \settowidth
	{\labelwidth}{#1.}\sloppy}}{\end{list}}
 
\newcounter{itemlistc}
\newcounter{romanlistc}
\newcounter{alphlistc}
\newcounter{arabiclistc}

\newcommand{\fcaption}[1]{
	\refstepcounter{figure}
	\setbox\@tempboxa = \hbox{\tenrm Fig.~\thefigure. #1}
	\ifdim \wd\@tempboxa > 6in
	   {\begin{center}
	\parbox{6in}{\tenrm\baselineskip=12pt Fig.~\thefigure. #1}
	    \end{center}}
	\else
	     {\begin{center}
	     {\tenrm Fig.~\thefigure. #1}
	      \end{center}}
	\fi}
 
\newcommand{\tcaption}[1]{
	\refstepcounter{table}
	\setbox\@tempboxa = \hbox{\tenrm Table~\thetable. #1}
	\ifdim \wd\@tempboxa > 6in
	   {\begin{center}
	\parbox{6in}{\tenrm\baselineskip=12pt Table~\thetable. #1}
	    \end{center}}
	\else
	     {\begin{center}
	     {\tenrm Table~\thetable. #1}
	      \end{center}}
	\fi}
 
\def\@citex[#1]#2{\if@filesw\immediate\write\@auxout
	{\string\citation{#2}}\fi
\def\@citea{}\@cite{\@for\@citeb:=#2\do
	{\@citea\def\@citea{,}\@ifundefined
	{b@\@citeb}{{\bf ?}\@warning
	{Citation `\@citeb' on page \thepage \space undefined}}
	{\csname b@\@citeb\endcsname}}}{#1}}
 
\newif\if@cghi
\def\cite{\@cghitrue\@ifnextchar [{\@tempswatrue
	\@citex}{\@tempswafalse\@citex[]}}
\def\citelow{\@cghifalse\@ifnextchar [{\@tempswatrue
	\@citex}{\@tempswafalse\@citex[]}}
\def\@cite#1#2{{$\null^{#1}$\if@tempswa\typeout
	{IJCGA warning: optional citation argument
	ignored: `#2'} \fi}}

\def\fnt#1#2{\footnotetext{\kern-.3em
	{$^{\mbox{\sevenrm #1}}$}{#2}}}
 
 1
 1
 1

\font\tenrm=cmr10

\font\ninerm=cmr9

\begin{document}
\begin{flushright}
FTUAM 95/21

FR-THEP-NR 97/9
\end{flushright}

\bigskip
\begin{center}

\large{WHY WE OBSERVE AN ALMOST CLASSICAL SPACETIME}
\end{center}
\begin{quote}
\begin{center}  
   J.L. Rosales \footnote{E-mail:rosales@phyq1.physik.uni-freiburg.de} 

   Fakult\"at f\"ur Physik, Universit\"at Freiburg,
   Hermann Herder Strasse 3, D-79104 Germany

   and

   J.L. S\'anchez-G\'omez \footnote{E-mail: jolu@vm1.sdi.uam.es}

    Departamento de F\'isica Te\'orica, Universidad Aut\'onoma de Madrid,
     Cantoblanco, E-28049-Madrid, Spain.
\end{center}
\end{quote}

\begin{quote}
\begin{center}
				Abstract
\end{center}
We argue  that, in order to obtain decoherence of spacetime, we should
consider quantum conformal metric fluctuations of spacetime.
This seems to be the  only required  envinonment
in the problem of selfmeasurement of spacetime in
quantum gravity formalism.

\end{quote}
\vspace{3 mm}

\section{Introduction}

It has been  suggested  (see e.g.,\cite{kn:Joos1}
and \cite{kn:Zeh1}) that it would be possible to demonstrate, by means
of a fully quantum
treatment, that spacetime becomes classical by a process
similar to that of the emergence of classical properties of a
macroscopic system in standard decoherence models. The latter arises
from  the quantum mechanical entanglement of the states of the
system with its environment: when introduced in the time evolution given
by the Schr\"odinger equation, the environment {\em measures} certain
properties
of the system thereby destroying the off-diagonal terms of the
density matrix in space representation \cite{kn:Joos2}.

On the other hand, the spacetime structure is classically obtained from
the matter distribution in the universe through Einstein's field
equations.
Nevertheless, the field equations do not tell us, by themselves, which
geometry is to be the geometry of spacetime, because they say nothing
{\em a priori} about the matter distribution of the universe ---Indeed,
they just state which geometries can
be possibly associated with a given matter distribution. We think this
observation to be in order
since it is a general belief that the environment in the quantum
mechanical treatment of the problem of {\em classicity} of spacetime structure
(i.e., the existence of correlations of the quantum states,
in the WKB sense, over the classical allowed trajectories)
should also involve some distribution of matter \cite{kn:Kiefer}-\cite{kn:Halliwell1}.
On the other hand, when spacetime is  non-classical, we do not
have necessarily to ``believe" in Einstein's field
equations \cite{kn:Page}, nor to think that the matter content of the
universe ``tells
space to behave classically" in Joos' words \cite{kn:Joos1}.

In this paper a new line of thought will be followed. As long as the
very nature
of spacetime should take into account, in quantum theory, the properties
of vacuum, \footnote {When dealing with the theory of gravity
there exists no vacuum at all, only {\em empty space solutions}.} then,
there should exist, in this framework,
a natural extension of what is considered as the environment. We think
that the {\em classicity}  of spacetime is a consequence of the
existence of some allowed
degrees of freedom in the evolution of the  manifold representing
the properties of gravitational empty solutions, which , without the
need of any particular
matter distribution, would lead to {\em classicity}.
As a simple model we propose the existence of conformal metric
fluctuations in the classical domain. The classical
allowed solutions have, of course, no dynamics at all
\footnote {if it were not so, then, the initial
value problem in general relativity would be physically inconsistent},
yet the corresponding quantum dynamics
will prove to be non trivial, giving rise to the expected
entanglement of the states of such fluctuations with the  non-quantized
variable of the cosmological model ---the scale factor.

We dedicate this paper to Prof. J.P. Vigier, whose pioneering work in
metric fluctuations has provided helpful suggestions to develop our
own approach.

\section{Classical dynamics of the conformal field}

From the hypothesis of the previous section, we are now going to develop
the dynamics of a model of conformal fluctuations in spacetime.
For the sake of mathematical simplicity, we will establish the hamiltonian
formulation of gravity  just in the isotropic cosmological model.

We are interested, for the time being, in the classical behaviour
of spacetime, then, in order to derive the dynamics,
we will use the Hilbert action corresponding to the following
metric
\begin{equation}
ds^{2}=l^{2}[N(t)^{2}dt^{2}-a(t)^{2}d\sigma^{2}] \mbox{,}
\end{equation}
where $d\sigma^{2}=\gamma_{ij}dx^{i}dx^{j}$ is the metric of the space-like
slices of the manifold and $l^{2}=2/3\pi m_{p}^{2}$

The Hilbert action (using natural units) is (see \cite{kn:DeWitt})

\begin{equation}
S=\frac{1}{16\pi}\int R(-g)^{1/2}dx^{4}=\frac{1}{2}\int(-\frac{a}{N}\dot{a}^{2}+Na)dt \mbox{.}
\end{equation}
where we have integrated the lagrangian density over the three spheres
of constant time.
Hence, we obtain a lagrangian functional for the metric variables
\begin{equation}
L=\frac{1}{2}(-\frac{a}{N}\dot{a}^{2}+Na) \mbox{.}
\end{equation}
Now consider the transformation
\begin{equation}
dt\rightarrow e^{\phi}dt=d\tilde{t}       \mbox{,}
\end{equation}
for some unspecified scalar function $\phi$. We are interested in those
scale functions $a(t)$ with the conformal transformation property given by
\begin{equation}
a(t)\rightarrow a(t)e^{\phi}=\tilde{a}(\tilde{t}) \mbox{.}
\end{equation}
This, of course, is equivalent to studying the dynamics of the conformal
functions $\phi$
\begin{equation}
d\tilde{s}^{2}=e^{2\phi}ds^{2}                        \mbox{.}
\end{equation}
Then the transformed metric reads

\begin{equation}
d\tilde{s}^{2}=l^{2}[\tilde{N}(\tilde{t})^{2}d\tilde{t}^{2}-\tilde{a}^{2}(\tilde{t})d\sigma^{2}] \mbox{,}
\end{equation}
where $\tilde{N}(\tilde{t})=N(t)$ and $\tilde{a}(\tilde{t})=e^{\phi}a(t)$.

But taking into account also the general covariance of the theory
and comparing (1) and (7), we obtain an identical expression for the transformed lagrangian in terms of its
new metric quantities
\begin{equation}
\tilde{L}=\frac{1}{2}(-\frac{\tilde{a}}{\tilde{N}}(\frac{d}{d\tilde{t}}{\tilde{a}})^{2}+\tilde{N}\tilde{a}) \mbox{.}
\end{equation}
Now, using the "old  coordinates"  we get, in terms of $\phi$
\begin{equation}
\tilde{L}=\frac{1}{2}(-\frac{e^{\phi}a(t)}{N(t)}e^{-2\phi}[\frac{d}{dt}(e^{\phi}a(t))]^{2}+a(t)N(t)e^{\phi}) \mbox{.}
\end{equation}

We could consider $N(t)$, $a(t)$ and $\dot{a}(t)$ as known
functions of time; therefore, as we are just interested in  the
classical dynamics which corresponds to the  $\phi$ field, we can define,
without any loss of generality, a particular set of functions. Let us consider
a time reparametrization such that $N(t)=1$. Upon doing this we finally get

\begin{equation}
\tilde{L_{a}}(\phi, \dot{\phi})=\frac{1}{2}a\{1-(\dot{\phi}a+\dot{a})^{2}\}e^{\phi} \mbox{.}
\end{equation}

We must now  develop the theoretical consequences involved in the
transformation
properties of this lagrangian. Thus, as we have defined $\phi$ to
be a generic
scalar function of time, we can explicitly state its transformation rule
under a time reparametrization, i.e.,
\begin{eqnarray}
a \rightarrow \tilde{a} &=& a+\alpha (a) \\ \nonumber
\phi(a)\rightarrow \tilde{\phi} & = &\phi+\delta \phi \\
\delta\phi & = & \dot{\phi}\alpha(a)
\end{eqnarray}
where we have taken into account that $\tilde{\phi}=\phi(\tilde{a})$.

In addition, the lagrangian transforms as a density, i.e.,
\begin{equation}
\tilde{L_{a}}(\tilde{\phi},\dot{\tilde{\phi}})=L_{a}(\phi,\dot{\phi})+\delta L
\end{equation}
where,
\begin{equation}
\delta L=\frac{d}{da}(L_{a}\alpha(a))
\end{equation}
On the other hand, using the transformation of $\phi$ (see (12))
\begin{eqnarray}
\delta L & = &\frac{\partial L_{a}}{\partial\phi}(\dot{\phi}\alpha(a))+
\frac{\partial L_{a}}{\partial\dot{\phi}}\frac{d}{da}(\dot{\phi}\alpha(a))+
\alpha(a)\frac{\partial L_{a}}{\partial a} =\\ \nonumber
& = &(\frac{\partial L_{a}}{\partial a}+\frac{\partial L_{a}}{\partial\phi} \dot{\phi}+
\frac{\partial L_{a}}{\partial\dot{\phi}}\ddot{\phi})\alpha(a)+
\frac{\partial L_{a}}{\partial\dot{\phi}}\dot{\phi}\dot{\alpha}(a)= \\ \nonumber
& = & \frac{d L_{a}}{da} \alpha(a)+\frac{\partial L_{a}}{\partial\dot{\phi}}\dot{\phi}\dot{\alpha}(a)= \\ \nonumber
& = & \frac{d (\alpha(a) L_{a})}{da} +(\frac{\partial L_{a}}{\partial\dot{\phi}}\dot{\phi}-L_{a})\dot{\alpha}(a) \mbox{,}
\end{eqnarray}
implying the typical hamiltonian constraint
\begin{equation}
H_{a}\equiv\frac{\partial L_{a}}{\partial\dot{\phi}}\dot{\phi}-L_{a}=0 \mbox{.}
\end{equation}

It is now straightforward to obtain this hamiltonian
function by means of the Legendre transformation (in terms of its coordinates
and canonical momenta)
\begin{equation}
H_{a}(\phi,p_{\phi})=p_{\phi}\dot{\phi}-L_{a}(\phi,\dot{\phi}\rightarrow p_{\phi}) \mbox{,}
\end{equation}
where
\begin{equation}
p_{\phi}=\frac{\partial L_{a}}{\partial\dot{\phi}}=- a^{2}e^{\phi}(\dot{a}+\dot{\phi}a) \mbox{,}
\end{equation}
which can also be inverted ($\dot{\phi}\rightarrow p_{\phi}$)
\begin{equation}
\dot{\phi}=-\frac{\dot{a}}{a}-\frac{1}{a^{3}}p_{\phi}e^{-\phi}   \mbox{.}
\end{equation}
Finally
\begin{equation}
H_{a}(\phi,p_{\phi})=-\{\frac{1}{2}a e^{\phi}+\frac{p_{\phi}\dot{a}}{a}+
\frac{p_{\phi}^{2}e^{-\phi}}{2 a^{3}}\}                          \mbox{.}
\end{equation}
Notice that we can cast this expression in a simpler, suggestive way
\begin{equation}
H_{a}(\phi,p_{\phi})=-[p_{\phi}-p_{0}(\phi,a^{2})]^{2}\frac{e^{-\phi}}{2 a^{3}}+(\dot{a}^{2}-1)\frac{e^{\phi} a}{2} \mbox{,}
\end{equation}
where
\begin{equation}
p_{0}(\phi,a^{2})=-\dot{a}a^{2}e^{\phi} \mbox{,}
\end{equation}
Taking into account the negativeness of the energy
for the gravitational field, we have to consider those cosmological
models satisfying  $\dot{a}^{2}\leq 1$. Therefore, upon assuming this,
the global negative sign of (21)  is a typical {\em footprint}
of the fact that $\phi$ is indeed a gravitational field.

For the sake of mathematical simplicity let us consider the particular ansatz
$\dot{a}(t)^{2}=1$, and $a(t)^{2}=t^{2}$; hence, we will identify cosmological time and
scale factor hereafter. Now, the constraint $H_{a}=0$ implies
\begin{equation}
[p_{\phi}-p_{0}(\phi,a^{2})]^{2}e^{-\phi}=0 \mbox{,}
\end{equation}
and, using again (18) ($p_{\phi}\rightarrow\dot{\phi}$) we
obtain the classical solutions of the conformal field
\begin{eqnarray}
p_{\phi} & = & p_{0}(\phi ,a)  \mbox{,}
\end{eqnarray}
or,
\begin{eqnarray*}
a^{2}e^{\phi}(1+\dot{\phi}a) & = & a^{2}e^{\phi} \mbox{,}
\end{eqnarray*}
that is $\dot{\phi}=0$.
But this is the condition required in the classical
theory since, in that case, $d\tilde{t}=e^{\phi}dt=d (e^{\phi} t)$. This
is now integrated to  get
$\tilde{t}=e^{\phi} t$, which is also the prescription for having
$\tilde{a}(\tilde{t})=e^{\phi}a(t)$ if and only if $a(t)=t$; indeed
this has been our choice of the scale factor function.

\section{Wheeler-DeWitt formalism in minisuperspace}

In spite of the trivial classical dynamics associated to
the conformal function $\phi$, equation (21) could be treated quantum
mechanically as a hamiltonian system. Thus, the quantum dynamics of this
system could, in principle, be expressed by means of the hamiltonian
constraint (16) together with the standard rule
for the canonical momentum in the operator formalism,
(i.e., $p_{\phi}\rightarrow \hat{p}_{\phi}=i\frac{\partial}{\partial \phi}$)
\begin{equation}
H_{a}(\phi,i\frac{\partial}{\partial \phi})\tilde{\Psi}(a^{2};\phi)=0 \mbox{.}
\end{equation}

The above equation is just the Wheeler-DeWitt equation
\begin{equation}
\{e^{-p\phi}[i\frac{\partial}{\partial\phi}+\dot{a}a^{2}e^{\phi}]^{2}e^{(p-1)\phi}+(\dot{a}^{2}-1)e^{\phi}a^{4}\}\tilde{\Psi}(a^{2};\phi)=0 \mbox{,}
\end{equation}
$p$ denotes the factor ordering ambiguity of the theory.

Notice that the effect of the background metric is just
a shift in the momentum of the conformal field. On the other hand, any
selection for the factor ordering  leads to a complex
time-dependent wave equation for pure
gravity
whose solutions (for $\dot{a}^{2}=1$) are given by
\begin{equation}
\tilde{\Psi}(a^{2};\phi)=\Psi(a)B_{p}(\phi)e^{i a^{2}e^{\phi}-1} \mbox{.}
\end{equation}
Here, $a$ is considered a {\em c-number},  $B_{p}(\phi)$ is a polynomial and
$\Psi(a)$ is again a constant with respect to the $\phi$-field; the latter
should be identified with the wave amplitude of the scale factor in its
configuration space (i.e., when it were not considered as a classical variable).
These wave functions have not a natural normalization.
On the other hand, if we  take $B_{p}=1$, then, we could obtain
a set of normalized wave functions in the sense of the Dirac delta
function.
To see this, we can study the classical equation corresponding to the
momentum constraint when $\dot{a}^{2}=1$ (see (24)). Upon making the standard replacement
$p_{\phi}\rightarrow \hat{p}_{\phi}=i\frac{\partial}{\partial \phi}$ we have
\begin{equation}
[i e^{-\beta\phi}\frac{\partial}{\partial \phi}e^{(\beta-1)\phi}+
a^{2}]\chi_{a^{2}}(\phi)=0 \mbox{,}
\end{equation}
$\chi_{a^{2}}(\phi)$ being the solutions of the momentum constraint and
$\beta$ representing again the uncertainty in the factor ordering.
Thus, $\beta=1$ is the natural choice since, in this case, we can
write the momentum constraint in terms of the relevant gravitational
quantity, i.e.,  the conformal field $\gamma\equiv e^{\phi}$; this
assumption being done,  (29) takes the simpler form
\begin{equation}
i\frac{\partial}{\partial \gamma}\chi_{a^{2}}(\gamma)=-a^{2}\chi_{a^{2}}(\gamma) \mbox{,}
\end{equation}
where we have put $e^{-\phi}\frac{\partial}{\partial \phi}=
\frac{\partial}{\partial \gamma}$.
But, as far as we are interested, in this particular case, in the continuum
spectrum of $\gamma$, we can try and normalize the wave functions using
the standard quantum mechanical prescription
\begin{equation}
\int \chi_{a^{2}}(\gamma)\chi^{*}_{\tilde{a}^{2}}(\gamma)d\gamma=\delta(a^{2}-\tilde{a}^{2} ) \mbox{,}
\end{equation}
or
\begin{equation}
\chi_{a^{2}}(\gamma)=(\frac{1}{2\pi})^{1/2}e^{i a^{2}\gamma} \mbox{.}
\end{equation}
Then, by using this eigenfunction basis we can also construct the
operator
whose eigenvalues coincide with the square of the scale factor values
corresponding to the background.
\begin{equation}
\hat{a_{\gamma}^{2}}\equiv -\hat{p}_{\gamma}=-[i\frac{\partial}{\partial \gamma}] \mbox{.}
\end{equation}
Hence, the eigenstate equation for this operator reads
\begin{equation}
\hat{a_{\gamma}^{2}}\chi_{a^{2}}(\gamma)=a^{2}\chi_{a^{2}}(\gamma) \mbox{.}
\end{equation}
We could then try and obtain the quantum evolution of the states
of the conformal fluctuations in terms of the cosmological time.

On the other hand, a general solution of (26) is given by (upon defining
$\gamma\equiv \frac{\dot{a}}{|\dot{a}|}e^{\phi}$)
\begin{eqnarray*}
\tilde{\Psi}_{\pm}(a;\gamma)=\tilde{\Psi}_{L}(a;\gamma) \pm \tilde{\Psi}_{R}(a;\gamma) \mbox{,}
\end{eqnarray*}
where the left ($\Psi_{L}(a;\gamma)$) and right ($\Psi_{R}(a;\gamma)$)
solutions of the Wheeler-DeWitt equation are given by
\begin{center}
\[\tilde{\Psi}_{L}(a;\gamma)= \left\{ \begin{array}{ll} =\Psi(a)e^{i|\dot{a}|a^{2}\gamma}e^{+a^{2}(1-\dot{a}^{2})^{1/2}(\gamma+1)} \mbox{ \hspace{1 mm} if \hspace{1 mm} $\gamma<-1$} & \\
=\Psi(a)e^{i|\dot{a}|a^{2}\gamma}e^{-a^{2}(1-\dot{a}^{2})^{1/2}(\gamma+1)} \mbox{ \hspace{1 mm} if \hspace{1 mm} $\gamma>-1$}  & \\
\end{array}
\right. \]
\end{center}
\begin{center}
\[\tilde{\Psi}_{R}(a;\gamma)= \left\{ \begin{array}{ll} =\Psi(a)e^{i|\dot{a}|a^{2}\gamma}e^{-a^{2}(1-\dot{a}^{2})^{1/2}(\gamma-1)} \mbox{ \hspace{1 mm} if \hspace{1 mm} $\gamma>1$} & \\
=\Psi(a)e^{i|\dot{a}|a^{2}\gamma}e^{+a^{2}(1-\dot{a}^{2})^{1/2}(\gamma-1)} \mbox{ \hspace{1 mm} if \hspace{1 mm} $\gamma< 1 $}  & \\
\end{array}
\right. \]
\end{center}

The wave functions are peaked about the classical allowed solutions, i.e.
  $\gamma^{2}=1$;
moreover, we can not single out any particular solution, thus leading to
interference between the expanding ($\gamma=+1$) and
collapsing ($\gamma=-1$) classical solutions (but for the case $\dot{a}^{2}=1$,
i.e., a matter free cosmolgical model where efective decoherence could be obtained
considering the conformal field as the environment).

\section{Time evolution equation}
Indeed, the problem of time in quantum gravity is that of giving sense
to the Wheeler-DeWitt formalism. This is so since, in the
context of quantum cosmology, waves have a trivial evolution.
It comes from the fact that in quantum gravity
the physical interest is represented by the scale factor itself (strictly
speaking three geometries and their configuration space, i.e.,
{\em superspace}) which could
be quantized in DeWitt sense of quantum gravity (see, for instance \cite{kn:DeWitt}
-\cite{kn:Rovelli}), i.e., the squared
of the wave function would lead to a probability for the quantized
metric of spacetime. Then, in dealing with this problem, there is no
evolution whatsoever; in fact, we are generally working in
minisuperspace, that is the quantization is often done  just for
the parametric time corresponding to the scale factor. Hence, time lies
out of the quantum formalism as a result of the hamiltonian constraint,
i.e., of the general covariance  of the theory.

Our problem is different. In spite of the fact that we have been dealing
with the parametric time, $a(t)$, we did not consider the possible
quantum states for $a(t)$ itself, i.e., we are not interested in defining a
probability for wave functions $\Psi(a)$. Yet, we are concerned in the problem
of the quantization of the conformal field (something analogous to the
{\em breathing modes} of the cosmological model). This led us to studying
wave functions defined, in principle, on the continuum spectrum; such
states were denoted in the previous section by $\chi_{a^{2}}(\gamma)$.
Moreover, $\{\chi_{a^{2}}(\gamma)\}$  could be used as a orthonormal
basis in order to develop the cosmological time evolution of generic operators,
for instance, that corresponding to the momentum ($p_{\gamma}$) or
the one of the scale factor squared ($\hat{a_{\gamma}^{2}}$, see  (32)
and (33)).To see this, we observe the cosmological time-reversal
invariance of our solutions in (31). This comes from the character of
a time-evolution wave equation which possesses $\chi_{a^{2}}(\gamma)$:
\begin{equation}
-i\frac{\partial}{\partial a}\Psi(a,\gamma)=2 a\gamma\Psi(a,\gamma) \mbox{.}
\end{equation}

Here we have defined $\Psi(a,\gamma)\equiv\chi_{a^{2}}(\gamma)$.

Now applying  the time reversal operator gives,
\begin{equation}
\hat{T}\{\Psi(a,\gamma)\}=\Psi(-a,\gamma) \mbox{,}
\end{equation}
while for the complex conjugate operator we obtain
\begin{equation}
\hat{C}\{\Psi(a,\gamma)\}=\Psi^{*}(a,\gamma)  \mbox{.} 
\end{equation}
Nevertheless, according to (34)
\begin{equation}
\hat{T}\{\Psi(a,\gamma)\}\neq \hat{C}\{\Psi(a,\gamma)\} \mbox{.}
\end{equation}
Then (34), though similar to  the Schr\"odinger equation, is not to be
regarded so in a very strict sense.

Now the physical meaningful cosmological solutions should not depend on
the environment. Moreover, we
have to take into account that we still do not know the properly normalized
solutions of the whole Wheeler-DeWitt operator  which, in general, should
depend on the cosmological model (i.e., it would develop other solutions when
the particular ansatz $\dot{a}^{2}=1$ were not made). In addition,
$\chi_{a^{2}}(\gamma)$ is just one
solution corresponding to the eigenstates of the momentum operator belonging to the kernel
of the hamiltonian constraint; this lack of information should be considered
in our model upon computing the reduced density matrix for the quantum states
of the envinonment;
the latter is being done upon tracing
out our solutions over the internal degrees of freedom for $\gamma$.
\begin{equation}
\tilde{\rho}(a^{2},\tilde{a}^{2})=\int_{-\infty}^{\infty}
\chi_{a^{2},\dot{a}}(\gamma)\chi_{\tilde{a}^{2},\dot{a}}^{*}(\gamma) D_{\dot{a}}\gamma \Psi(a)\Psi^{*}(\tilde{a}) \mbox{.}
\end{equation}

Here,$D_{\dot{a}}\gamma$ denotes a measure corresponding to the environment when
a general solution is considered.

\section{Quantum Correlations}

We have obtained a time evolution equation for the  states of the quantum
conformal metric fluctuation corresponding to an isotropic background
gravitational field in terms of the cosmological time $a$. Then, in order
to obtain a generalization of the Schr\"odinger formalism, we could try
developing the initial value problem for this equation.

Yet, as long as we do not know the internal freedom for $\gamma$,
we should model this ambiguity upon asuming the initial state being
correlated with the classical solution (i.e., $\gamma=1$).
Hence,let us
consider an initial wave packet given, in terms of the $\gamma$-field, by
\begin{equation}
\Psi(0, \gamma)=\sigma^{-1/2}\pi^{-1/4}e^{\frac{(\gamma-1)^{2}}{\sigma^{2}}} \mbox{,}
\end{equation}
where $\sigma$ is a constant which characterizes
the minimal uncertainty of the conformal field $\gamma$.
Then, at a different time, say $a$, by Fourier-transforming we get
from (34)
\begin{equation}
\Psi(a,p_{\gamma})=\frac{1}{(2\pi)^{1/2}}\int_{-\infty}^{+\infty}
e^{i(p_{\gamma}\gamma + a^{2}(\gamma-1)+ a^{2})}
\cdot \Psi(0,\gamma) d\gamma \mbox{,}
\end{equation}
or
\begin{equation}
\Psi(a,p_{\gamma})=(\frac{\sigma}{\pi^{1/2}})^{1/2}
e^{i(a^{2} +p_{\gamma})}
e^{-\frac{\sigma^{2}}{2}(p_{\gamma}+ a^{2})^{2}} \mbox{.}
\end{equation}

Now, following the usual prescription for the probability amplitude,
we get
\begin{equation}
\rho_{\sigma}(a,p_{\gamma})=|\Psi(a,p_{\gamma})|^{2}=\frac{\sigma}{\pi^{1/2}}
e^{-\sigma^{2}(p_{\gamma}+ a^{2})^{2}}           \mbox{.}
\end{equation}

On the other hand, we can also obtain a probability measure not only for
the momentum $p_{\gamma}$ but also for any function $F(p_{\gamma})$;
for instance, we may obtain the probability measure for the operator
corresponding to the scale factor squared (see (32)),
\begin{equation}
\rho_{\sigma}(a,\tilde{a}^{2})=|\frac{\partial p_{\gamma}}{\partial a^{2}}|
\rho_{\sigma}(a,p_{\gamma}\rightarrow \tilde{a}^{2}) \mbox{,}
\end{equation}
where we have made use of the jacobian of the transformation. That is
\begin{equation}
\rho_{\sigma}(a,\tilde{a}^{2})
=\frac{\sigma}{\pi^{1/2}}e^{-\sigma^{2}(a^{2}-\tilde{a}^{2})^{2}} \mbox{.}
\end{equation}
Here, we have denoted by $\tilde{a}^{2}$ the eigenvalues of $\hat{a^{2}}$.

On the other hand, if we take the limit $\sigma\rightarrow\infty$ in (45),
\begin{eqnarray}
\rho(a,\tilde{a}^{2})=\lim_{\sigma\rightarrow\infty}\rho_{\sigma}(a,\tilde{a}^{2})
& = & \lim_{\sigma\rightarrow\infty}\frac{\sigma}{\pi^{1/2}}e^{-\sigma^{2}(a^{2}-\tilde{a}^{2})^{2}}=\delta(a^{2}-\tilde{a}^{2}) \mbox{.}\\ \nonumber 
\end{eqnarray}
Therefore, in order to achieve complete decoherence of space-time we
required some  uncorrelated initial state for the conformal degree
of freedom of the metric.

\section{Discussion}
Let us now try and put the above result into context. We have seen that
the almost classical  character of spacetime could be a consequence of the initial
conditions of conformal metric fluctuations. This restricts the set of
quantum states to the {\em expanding} (forward time solution) and {\em collapsing} ones
(backward time solution), therefore, there would
exist possible tunneling between both physical allowed states, i.e.,
{\em interference}. Moreover, the
Weyl tensor is  the only geometrical invariant under the conformal field;
it implies that since upon considering conformal type fluctuations,
we have obtained two possible quantum states, then
the way spacetime would become classical (i.e., the selection of one of these
two branches of the solution) should only be
a consequence of the initial conditions in this quantity.
On the other hand, this tensor is precisely that part of the Riemannian
curvature which is source free, a fact that strengthens our previous
believe that classical properties of spacetime should not be a consequence
of any particular model corresponding to a matter field environment
(It is also somehow in accordance with Penrose proposal
\cite{kn:Penrose}.)

We have seen that spacetime may become classical when the density matrix
(see (45)) is peaked about the classical allowed configurations.
Moreover, the feature
of exponential behaviour depends on the initial conditions for the
environment (in order to see this, recall that the decoherence
width of the geometrical
scale parameter satisfies $\delta a^{2}\sim \sigma^{-1}$). The latter
agrees with Hartle's conjecture \cite{kn:Hartle1}, \cite{kn:Hartle2}
(see also \cite{kn:Halliwell2}); from this point of view the initial
conditions control the extent to which macroscopic states decohere.

A remarkable fact is that the scale factor becomes sharply peaked
about the classical solution when  the limit
$\sigma\rightarrow\infty$ is considered, that is, when the initial quantum fluctuations of
spacetime are very large; it seems to be a typical behaviour of
a phase transition process
(i.e.,large scale fluctuations in a system also develop
far from the equilibrium correlations).

\end{document}